\documentclass[a4paper,preprintnumbers,aps,prc]{revtex4}
\usepackage{latexsym}
\usepackage{graphics}
\usepackage{amsmath}
\usepackage{amssymb}
\begin{document}
\title{Extended Goldstone-Boson-Exchange Constituent Quark Model}
\author{K. Glantschnig, R. Kainhofer, W. Plessas, B. Sengl, and R.F. Wagenbrunn}

%
%
\affiliation{Theoretical Physics, Institute of Physics, University of Graz, 
Universit\"atsplatz 5, A-8010 Graz, Austria}
%
%
\begin{abstract}
We present an extension of the Goldstone-boson-exchange constituent
quark model including additional interactions beyond the ones used 
hitherto. For the hyperfine interaction 
between the constituent quarks we assume pseudoscalar, vector, and scalar 
meson exchanges and consider all relevant force 
components produced by these types of exchanges. The resulting model, 
which corresponds to a relativistic Poincar\'e-invariant Hamiltonian 
(or equivalently mass operator), provides a unified framework for a
covariant description of all light and strange baryons. The ground
states and resonances up to an excitation energy of about 2 GeV are 
reproduced in fair agreement with phenomenology, with the exception of 
the first excitations above the $\Lambda$ and $\Xi$ ground states.
\end{abstract}
%
%
\maketitle

\section{Introduction}
\label{intro}

It is nowadays widely accepted that quantum chromodynamics (QCD) is the 
basic theory of strong interactions. Over the years, however, one has
learned that QCD manifests itself with different degrees of freedom in 
different energy regimes. As a consequence one resorts to separate 
approaches for the solution of QCD depending on the energy domain one 
is working in. Low-energy QCD is characterized by the appearance of
constituent quarks as quasiparticles (see, e.g., the lattice studies
in ref. \cite{Aoki99}). Therefore, it seems
to be appropriate to treat hadrons in terms of constituent quarks. 
An effective tool towards a comprehensive description of low-energy 
hadron phenomena is offered by constituent quark models (CQMs).
They can be designed to incorporate the essential ingredients of 
low-energy QCD. At the same time they can make use of a
covariant formulation so as to allow for the necessary relativistic 
treatment of hadrons as composite systems of constituent quarks
confined to a relatively small volume. When building a CQM one 
faces the problem of the proper effective interaction between the 
constituent quarks. While the confinement can readily be modeled after 
lattice results of QCD (see, e.g., ref. \cite{Bali:2000}), there are 
alternative approaches to the hyperfine interaction of the 
constituent quarks.

One type of hyperfine interaction between constituent quarks derives 
from one-gluon exchange (OGE) \cite{RGG:75}. Several CQMs are
based on this dynamical concept \cite{IK:78,CKP:83,SG:85,GI:85,CI:86}.
While OGE CQMs have a long tradition, all of them are facing some
serious problems especially in baryon spectroscopy. In particular, 
they do not manage to describe in a uniform manner the level schemes 
in the $N$, $\Delta$, and $\Lambda$ spectra in accordance with 
experimental data \cite{GPPVW:98}.

Another approach to the hyperfine interaction of constituent quarks 
considers instanton-induced effects \cite{LKMP:01,LMP:01,LMPe:01}. The 
CQM based on this concept also 
encounters similar difficulties in the $N$ and $\Delta$ spectra as it 
cannot reproduce the correct level orderings of positive- and 
negative-parity excitations \cite{Plessas03}.

Congruent with the assumption of constituent quarks as the essential 
degrees of freedom at low energies is the appearance of Goldstone 
bosons. Therefore, it seems to be quite natural to derive the hyperfine 
interaction from Goldstone-boson exchange (GBE) \cite{glozrisk}.
A CQM based on this 
dynamical concept has indeed been quite successful in describing the
excitation spectra of the light and strange baryons \cite{GPVW:98}.
The specific 
spin-flavor symmetry that is brought about by GBE allows to reproduce 
in a unified framework specifically the characteristic patterns in 
the level schemes of the $N$, $\Delta$, and $\Lambda$ spectra
\cite{GPPVW:98,GPVW:98}.

The CQM of ref. \cite{GPVW:98} uses only a part of the dynamics offered 
by GBE, namely, only the spin-spin part of the pseudoscalar 
exchange. Obviously, one is interested in the performance of a more 
complete model that takes into account also the other force 
components of the single Goldstone-boson exchange, such as the 
complete tensor force, as well as the possibilities offered by
multiple GBE \cite{RB2000}. First preliminary attempts in this
direction were already made in refs.
\cite{WPGV:99,WPGV:99a,WGPV:00,PGVW:00a}. In the present
paper we report an extended version of the GBE CQM that includes all
relevant force components offered by pseudoscalar, vector,
and scalar exchanges.

We also address the role of spin-orbit forces in more detail. 
Spin-orbit forces can stem from both the confinement and the hyperfine
interactions. From phenomenology, however, one expects their net
effect to be small in baryon spectroscopy. The more this should be 
true for quadratic spin-orbit forces. Therefore, in CQMs
it is often assumed that spin-orbit as well as quadratic spin-orbit
forces can be left out. Nevertheless, we study 
the influences of spin-orbit forces by considering two versions
of the extended GBE CQM, one without and one with such type of 
interaction. 

This work is organized as follows: In section \ref{sec:ps_GBE_CQM} we 
shortly recapitulate the properties of the pseudoscalar GBE 
CQM of ref. \cite{GPVW:98}, which is restricted to spin-spin forces 
of pseudoscalar exchange only.
In section \ref{sec:multiple_GBE} we discuss the idea of multiple 
GBE and its effective description.
The formulation of the extended GBE CQM is presented in section
\ref{sec:extended_GBE}. In section \ref{sec:parameterization} we 
give the parameterizations of the two versions of the extended
GBE CQM constructed here. The resulting spectra of
all the light and strange baryons are presented in section
\ref{sec:results}. We close with a discussion of the obtained results 
and an outlook to possible applications and tests of the new type of 
GBE CQM in low-energy hadronic physics.

\section{Pseudoscalar Goldstone-Boson-Exchange Constituent Quark Model}
\label{sec:ps_GBE_CQM}

In ref. \cite{GPVW:98} one constructed a CQM relying on the 
Hamiltonian
\begin{equation}
H=\sum^3_{i=1}\sqrt{\vec{p}^2_i+m^2_i}+\sum_{i<j}[V_{conf}(i,j)+V_{hf}(i,j)] ,
\label{ps_ham}
\end{equation}
where $\vec{p}_i$ are the three-momenta of the constituent quarks and $m_i$
are their masses. The confinement potential $V_{conf}$ was chosen to be of 
a linear form, and the hyperfine potential $V_{hf}$ consisted of the 
spin-spin components of the pseudoscalar meson exchange ($\pi, K, \eta, 
\eta '$). The Hamiltonian (\ref{ps_ham}) corresponds to an invariant 
(interacting) mass operator $M$ in relativistic quantum mechanics and 
it lends itself to covariant calculations not only of the energy spectra
but also of further hadronic reactions. 

The spectra and wave functions of the light and strange baryons were 
obtained by solving the eigenvalue problem of the above Hamiltonian 
(or equivalently of the mass operator $M$) via the 
stochastic variational method (SVM) \cite{VS:95,SV:98}.
This approach provides a very reliable determination of the 
eigenenergies and eigenstates. The accuracy of the method was checked
before by solving the problem of a relativistic three-quark system
via (modified) Faddeev equations \cite{KPP}.

\begin{figure}
\begin{center}
\resizebox{0.4\textwidth}{!}{\includegraphics{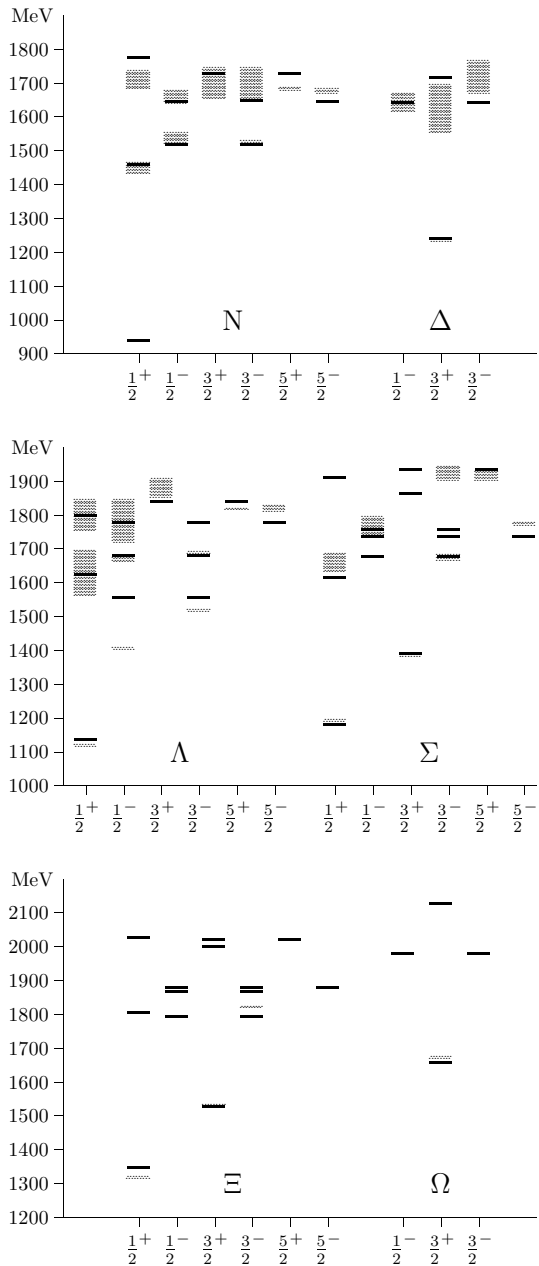}}
\end{center}
\caption{Energy levels of the lowest light and strange baryon states for 
the GBE CQM of ref. \cite{GPVW:98}. The mass of the
nucleon ground state is 939 MeV. The shadowed boxes represent the 
experimental values together with their uncertainties \cite{PDG}.}
\label{fig:ps_gbe}
\end{figure}

The spectra for the light and strange baryons resulting from the 
pseudoscalar GBE CQM are shown in figure \ref{fig:ps_gbe}.
As is immediately evident, quite a satisfactory description of the 
spectroscopy
of all light and strange baryons can be achieved already with this 
kind of GBE CQM. Especially, the orderings of the 
energy levels according to their parities are obtained in agreement with 
experiment. In the spectrum of the nucleon, the Roper resonance $N$(1440) 
with $J^P=\frac{1}{2}^+$ comes out as the lowest excitation above the 
nucleon ground state. It is followed by the $\frac{1}{2}^-- \frac{3}{2}^-$ 
doublet $N$(1535)--$N$(1520). In the spectra of the  $\Lambda$ and the $\Sigma$
the $\frac{1}{2}^+$ resonances $\Lambda (1600)$ and $\Sigma (1660)$ fall 
below the resonances with negative parity $\Lambda (1670)$, $\Lambda (1690)$, 
and $\Sigma (1750)$. In addition, the two $\frac{1}{2}^--\frac{3}{2}^-$ 
states $\Lambda (1405)$--$\Lambda (1520)$ remain the lowest excitations 
above the ground state. This behavior rests on the explicit flavor dependence 
of the GBE CQM and is governed by the 
particular spin-flavor symmetry coming with GBE.

With the specific type of hyperfine interaction deriving from GBE the 
intricate problem of the correct ordering
of the low-lying excitations of the light and strange baryons
is thus readily resolved. However, there remain some other problems.
Most strikingly the $\Lambda(1405)$ resonance remains far from the 
experimental value. Furthermore, some states in the bands of higher 
excitations in the $N$ (as well as the $\Lambda$) spectrum do not fit 
the experimental data; especially the near degeneracies of the
$\frac{1}{2}^+-\frac{1}{2}^-$ and similarly of the
$\frac{3}{2}^+-\frac{3}{2}^-$ and $\frac{5}{2}^+-\frac{5}{2}^-$ $N$
resonances at about 1700 MeV are not reproduced.

In this situation an obvious question was if further
improvements of the description of the light and strange baryon spectra
could be obtained by an extension of the pseudoscalar GBE CQM 
to including further force components. Wagenbrunn et al. made first
attempts to extend the GBE CQM and
proceeded to study the influences of the tensor
forces from pseudoscalar GBE \cite{WPGV:99,WPGV:99a}. However, with 
these tensor forces alone no improvement could be achieved, because 
the remarkably small splittings in the various multiplets of alike
total-angular-momentum states could not be reproduced as observed in 
experiment. From phenomenology one expects the total tensor-force 
effects to be small. Therefore one went ahead to include also vector- 
and scalar-type exchanges in addition to the pseudoscalar one
\cite{WPGV:99}. The spin-orbit forces, however, had then not yet been
taken into account in the parametrization of an extended GBE CQM
\cite{WGPV:00,PGVW:00a}.
\section{Multiple Goldstone-Boson Exchange}
\label{sec:multiple_GBE}

\begin{figure}[ht]
\begin{center}
\resizebox{0.45\textwidth}{!}{\includegraphics{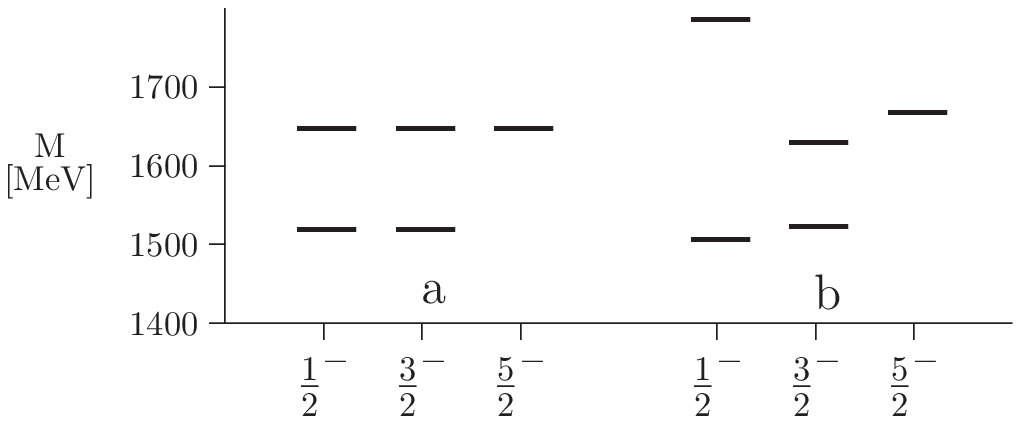}}
\end{center}
\caption{LS-multiplets of nucleon excitation levels for the 
pseudoscalar GBE CQM: a) without tensor forces (as in fig. 
\ref{fig:ps_gbe}), b) with tensor forces.}
\label{lsmulti}
\end{figure}

In the pseudoscalar GBE CQM a reasonable description of the light and 
strange baryon spectra is achieved only if the hyperfine 
interaction is based solely on the spin-spin part. If the tensor 
forces are included (with a strength as prescribed by unique 
octet and singlet pseudoscalar coupling constants, with values
${g^2_{ps,8}}/{4\pi}=0.67$ and ${g^2_{ps,0}}/{4\pi}=0.9$, 
respectively, see ref. \cite{GPVW:98}), one can no longer
keep certain splittings small, as demanded by phenomenology. This
is true especially for the
$\frac{1}{2}^--\frac{3}{2}^-$ resonances at around 1510 MeV as well as 
the $\frac{1}{2}^--\frac{3}{2}^--\frac{5}{2}^-$ resonances at around
1650 MeV. The behavior
is demonstrated in figure \ref{lsmulti}, where in panel b 
the effects of the tensor forces due to pseudoscalar meson exchange 
are demonstrated; obviously the generated splittings are too large 
especially in the second band. 

In the picture of GBE dynamics there is also the possibility of 
multiple-boson exchanges, in addition to the single-boson exchange 
prevailing in the ansatz of pseudoscalar coupling \cite{RB2000}. One
may take the effects of such multiple-boson exchanges into account
by foreseeing single-boson exchanges of scalar and vector type,
just as in the meson-exchange model of the nucleon-nucleon interaction
\cite{MHE:87,M:89}. For example, the effects of two-pion exchanges can 
effectively be described by $\sigma$-meson exchange ($\pi\pi$ S-wave)
and by $\rho$-meson exchange ($\pi\pi$ P-wave),
the effects of three-pion exchanges by $\omega$-meson
exchange, and so forth. 
One may assume that an analogous approach works in the case of 
Goldstone-boson-exchange dynamics between constituent quarks.

\section{Extended Goldstone-Boson-Exchange Constituent Quark Model}
\label{sec:extended_GBE}

Motivated by the discussion above we constructed our extended version of 
the GBE CQM. For this purpose we considered pseudoscalar, vector, and scalar 
meson exchanges, as well as all relevant force components (i.e. central, 
spin-spin, tensor, and spin-orbit forces). The Hamiltonian of the extended 
GBE CQM has the following form
\begin{equation}
H=\sum^3_{i=1}\sqrt{\vec{p}^2_i+m^2_i}+\sum_{i<j}[V_{conf}(i,j)+V_{hf}(i,j)].
\label{ext_ham}
\end{equation}
Again $\vec{p}_i$ are the three-momenta and $m_i$ are the masses of the 
constituent quarks. $V_{conf}(i,j)$ is the confinement potential and 
$V_{hf}(i,j)$ denotes the hyperfine interaction between quark $i$ and $j$. 
It is now taken to be the sum of the pseudoscalar (ps), vector (v), and 
scalar (s) meson exchange potentials,
\begin{eqnarray}
V_{hf}(i,j)&=&V^{ps}(i,j)+V^{v}(i,j)+V^{s}(i,j)\nonumber\\            
&=&\sum^3_{a=1}\left[V_{\pi}(i,j)+V_{\rho}(i,j)+V_{a_0}(i,j)\right]\lambda^a_i\cdot
\lambda^a_j+\nonumber\\
&+&\sum^7_{a=4}\left[V_{K}(i,j)+V_{K^*}(i,j)+V_{\kappa}(i,j)\right]\lambda^a_i\cdot
\lambda^a_j+\nonumber\\
&+&\left[V_{\eta}(i,j)+V_{\omega_8}(i,j)+V_{f_0}(i,j)\right]\lambda^8_i\cdot\lambda^8_j+
\nonumber\\
&+&\frac{2}{3}\left[V_{\eta'}(i,j)+V_{\omega_0}(i,j)+V_{\sigma}(i,j)\right] ,
\label{fullpot}
\end{eqnarray}
where $\lambda_i^a$ denote the Gell-Mann flavor matrices.

Subsequently we shall give the explicit expressions for the 
phenomenological confinement potential as well as for the spatial parts of 
the different meson-exchange potentials using the following form factor for
the constituent quark-meson vertex \cite{PGVW:00a}
\begin{equation}
F\left(\vec{q}\thinspace^2\right)=\frac{\Lambda^2_{\gamma}-\mu^2_{\gamma}}
{\Lambda^2_{\gamma}+\vec{q}\thinspace^2}.
\label{form_factor}
\end{equation}
Here, $\vec{q}$ is the three-momentum of the exchanged meson $\gamma$, 
$\mu_{\gamma}$ its mass, and $\Lambda_{\gamma}$ the corresponding cut-off 
parameter. For mesons with different masses these cut-off parameters may be 
different. Their values should be bigger for mesons with larger masses.

\subsection{Confinement}

For the confinement interaction we assume a linear potential dependent 
on the distance $r_{ij}$ between constituent quarks $i$ and $j$
\begin{equation}
V_{conf}(i,j)=V_0+C r_{ij}.
\label{confinement}
\end{equation}
The strength $C$ is treated as a fit parameter but it turns out to be 
of a magnitude roughly corresponding to the string tension of QCD. The
constant $V_0$ is needed to fix the ground-state energy of the 
spectrum, i.e. the nucleon, to its phenomenological value of 939 
MeV.

According to lattice QCD calculations \cite{Bali:2000} one would 
expect an additional (short-range) Coulomb term of the form $-a/r_{ij}$ 
(with strength $a$) in the confinement potential.
Our studies, however, have revealed that such a term 
can effectively be absorbed into the remaining potential parts. In particular,
its effect can be compensated by a slight adjustment of the linear term
$C r_{ij}$ and a small variation of the cut-off parameter
$\Lambda_{\sigma}$ in the scalar exchange potential \cite{Ka:03}. Thus,
in all our calculations we just employ the linear confinement potential
\eqref{confinement}.

\subsection{Pseudoscalar Part}
The pseudoscalar meson-exchange interaction ($\gamma=$ $\pi,$ $K,$ $\eta,$ 
$\eta '$) produces spin-spin and tensor forces. The corresponding 
potential is
\begin{eqnarray}
V_{\gamma}\left(i,j\right)&=&V^{SS}_{\gamma}\left(\vec{r}_{ij}\right)\vec
{\sigma}_i\cdot\vec{\sigma}_j\nonumber\\[4pt]
&+&V^{T}_{\gamma}\left(\vec{r}_{ij}\right)\left[3\left(\hat{r}_{ij}\cdot
\vec{\sigma}_i\right)\left(\hat{r}_{ij}\cdot\vec{\sigma}_j\right)-\vec
{\sigma}_i\cdot\vec{\sigma}_j\right],
\label{ps_pot}
\end{eqnarray}
with $V^{SS}$ the spin-spin and $V^T$ the tensor components; 
$\vec{\sigma}_i$ denotes the spin operator of quark $i$. 
The dependences of $V^{SS}$ and $V^T$ on the interquark separation 
$\vec{r}_{ij}=\vec{r}$ are given as follows: \vspace{3mm} \newline
Spin-spin component:
\begin{eqnarray}
V^{SS}_{\gamma}\left(\vec{r}\!\right)&=&\frac{g^2_{\gamma}}{4\pi}\frac{1}
{12m_im_j}\left[\mu^2_{\gamma}\frac{e^{-\mu_{\gamma}r}}{r}-\qquad\qquad
\right.\nonumber\\[4pt]
&&\qquad\qquad-\left.\left(\mu^2_{\gamma}+\frac{\Lambda_{\gamma}\left(
\Lambda^2_{\gamma}-\mu^2_{\gamma}\right)r}{2}\right)\frac{e^{-\Lambda_{
\gamma}r}}{r}\right];\nonumber\\
\label{ps2}
\end{eqnarray}
Tensor component:
\begin{eqnarray}
V^{T}_{\gamma}\left(\vec{r}\!\right)&=&\frac{g^2_{\gamma}}{4\pi}\frac{1}
{12m_im_j}\left[\mu^2_{\gamma}\left(1+\frac{3}{\mu_{\gamma}r}+\frac{3}
{\mu^2_{\gamma}r^2}\right)\frac{e^{-\mu_{\gamma}r}}{r}\right.-\nonumber\\
[4pt]
&&\qquad\qquad\quad-\Lambda^2_{\gamma}\left(1+\frac{3}{\Lambda_{\gamma}r}+
\frac{3}{\Lambda^2_{\gamma}r^2}\right)\frac{e^{-\Lambda_{\gamma}r}}{r}- 
\nonumber\\[4pt]
&&\qquad\qquad\quad-\left.\frac{\left(\Lambda^2_{\gamma}-\mu^2_{\gamma}
\right)\left(1+\Lambda_{\gamma}r\right)}{2}\frac{e^{-\Lambda_{\gamma}r}}
{r}\right].\nonumber\\
\label{ps3}
\end{eqnarray}
Here, $g_{\gamma}$ represents the quark-meson coupling constant. Again,
$m_{i}$ are the constituent quark masses, $\mu_{\gamma}$ the meson masses,
and $\Lambda_{\gamma}$ the corresponding cut offs. 

\subsection{Vector Part}
The vector meson-exchange interaction ($\gamma=$ $\rho,$ $K^*,$ $\omega_8,$ 
$\omega_0$) produces central, spin-spin, tensor, as well as spin-orbit forces. 
The corresponding potential is
\begin{eqnarray}
V_{\gamma}\left(i,j\right)&=&V^{C}_{\gamma}\left(\vec{r}_{ij}\right)+V^{SS}
_{\gamma}\left(\vec{r}_{ij}\right)\vec{\sigma}_i\cdot\vec{\sigma}_j\nonumber
\\[4pt]
&+&V^{T}_{\gamma}\left(\vec{r}_{ij}\right)\left[3\left(\hat{r}_{ij}\cdot
\vec{\sigma}_i\right)\left(\hat{r}_{ij}\cdot\vec{\sigma}_j\right)-\vec{
\sigma}_i\cdot\vec{\sigma}_j\right]\nonumber\\[4pt]
&+&V^{LS}_{\gamma}\left(\vec{r}_{ij}\right)\vec{L}_{ij}\cdot\vec{S}_{ij},
\label{v1}
\end{eqnarray}
where $\vec{L}_{ij}$ and $\vec{S}_{ij}$ are the two-body 
angular-momentum and spin operators. The dependences of $V^C$, $V^{SS}$,
$V^T$, and $V^{LS}$ on the interquark separation 
$\vec{r}_{ij}=\vec{r}$ are given as follows: \vspace{3mm} \newline
Central component:
\begin{eqnarray}
V^{C}_{\gamma}\left(\vec{r}\!\right)&=&\frac{\left(g^V_{\gamma}\right)^2}
{4\pi}\left[\frac{e^{-\mu_{\gamma}r}}{r}-\left(1+\frac{\left(\Lambda^2_{
\gamma}-\mu^2_{\gamma}\right)r}{2\Lambda_{\gamma}}\right)\frac{e^{-\Lambda
_{\gamma}r}}{r}\right];\nonumber\\
\label{v2}
\end{eqnarray}
Spin-spin component:
\begin{eqnarray}
V^{SS}_{\gamma}\left(\vec{r}\!\right)&=&2\frac{\left(g^V_{\gamma}+g^T_{
\gamma}\right)^2}{4\pi}\frac{1}{12m_im_j}\left[\mu^2_{\gamma}\frac{e^{-
\mu_{\gamma}r}}{r}-\qquad\qquad\right.\nonumber\\[4pt]
&&\qquad\qquad-\left.\left(\mu^2_{\gamma}+\frac{\Lambda_{\gamma}\left(
\Lambda^2_{\gamma}-\mu^2_{\gamma}\right)r}{2}\right)\frac{e^{-\Lambda_{
\gamma}r}}{r}\right];\nonumber\\
\label{v3}
\end{eqnarray}
Tensor component:
\begin{eqnarray}
V^{T}_{\gamma}\left(\vec{r}\!\right)&=&-\frac{\left(g^V_{\gamma}+g^T_{
\gamma}\right)^2}{4\pi}\frac{1}{12m_im_j}\times\qquad\qquad\nonumber\\[4pt]
&&\qquad\qquad\times\left[\mu^2_{\gamma}\left(1+\frac{3}{\mu_{\gamma}r}+
\frac{3}{\mu^2_{\gamma}r^2}\right)\frac{e^{-\mu_{\gamma}r}}{r}\right.- 
\nonumber\\[4pt]
&&\qquad\qquad-\Lambda^2_{\gamma}\left(1+\frac{3}{\Lambda_{\gamma}r}+\frac
{3}{\Lambda^2_{\gamma}r^2}\right)\frac{e^{-\Lambda_{\gamma}r}}{r}- \nonumber
\\[4pt]
&&\qquad\qquad-\left.\frac{\left(\Lambda^2_{\gamma}-\mu^2_{\gamma}\right)
\left(1+\Lambda_{\gamma}r\right)}{2}\frac{e^{-\Lambda_{\gamma}r}}{r}\right];\quad
\label{v4}
\end{eqnarray}
Spin-orbit component:
\begin{eqnarray}
V^{LS}_{\gamma}\left(\vec{r}\!\right)&=&-\frac{\left(g^V_{\gamma}\right)^2}{4\pi}\left(3+4\frac{g^T_{\gamma}}{g^V_{\gamma}}\right)\frac{1}{2m_im_j}\times\qquad\qquad\nonumber\\[4pt]
&\times&\left[\mu^3_{\gamma}\left(\frac{1}{\mu^2_{\gamma}r^2}+\frac{1}
{\mu^3_{\gamma}r^3}\right)e^{-\mu_{\gamma}r}\right.- \nonumber\\[4pt]
&-&\Lambda^3_{\gamma}\left(\frac{1}{\Lambda^2_{\gamma}r^2}+\frac{1}
{\Lambda^3_{\gamma}r^3}\right)e^{-\Lambda_{\gamma}r}-\left.\frac{\Lambda^2_
{\gamma}-\mu^2_{\gamma}}{2r}e^{-\Lambda_{\gamma}r}\right].\nonumber\\[4pt]
\label{v5}
\end{eqnarray}
The notation is the same as before, only we encounter different vector and 
tensor coupling constants $g^V_{\gamma}$ and $g^T_{\gamma}$, 
respectively.

In principle, the vector meson exchange (like the subsequent scalar 
meson exchange) also produces a quadratic spin-orbit interaction. 
Since it is of higher order in the inverse quark masses, it is 
expected to be of minor importance. Therefore it is neglected here 
(and below in the scalar meson exchange).

\subsection{Scalar Part}
The scalar meson-exchange interaction ($\gamma=$ $a_0,$ $\kappa,$ $f_0,$ 
$\sigma$) produces only central and spin-orbit forces. The corresponding
potential is
\begin{equation}
V_{\gamma}\left(i,j\right)=V^{C}_{\gamma}\left(\vec{r}_{ij}\right)+V^{LS}_
{\gamma}\left(\vec{r}_{ij}\right)\vec{L}_{ij}\cdot\vec{S}_{ij}.
\label{s1}
\end{equation}
The dependences of $V^C$ and $V^{LS}$ on the interquark separation 
$\vec{r}_{ij}=\vec{r}$ are given as follows: \vspace{3mm} \newline
Central component:
\begin{eqnarray}
V^{C}_{\gamma}\left(\vec{r}\!\right)&=&-\frac{g_{\gamma}^2}{4\pi}\left[
\frac{e^{-\mu_{\gamma}r}}{r}-\left(1+\frac{\left(\Lambda^2_{\gamma}-\mu^2
_{\gamma}\right)r}{2\Lambda_{\gamma}}\right)\frac{e^{-\Lambda_{\gamma}r}}
{r}\right];\nonumber\\
\label{s2}
\end{eqnarray}
Spin-orbit component:
\begin{eqnarray}
V^{LS}_{\gamma}\left(\vec{r}\!\right)&=&-\frac{g_{\gamma}^2}{4\pi}\frac{1}
{2m_im_j}\left[\mu^3_{\gamma}\left(\frac{1}{\mu^2_{\gamma}r^2}+\frac{1}{
\mu^3_{\gamma}r^3}\right)e^{-\mu_{\gamma}r}\right.- \nonumber\\[4pt]
&-&\Lambda^3_{\gamma}\left(\frac{1}{\Lambda^2_{\gamma}r^2}+\frac{1}{
\Lambda^3_{\gamma}r^3}\right)e^{-\Lambda_{\gamma}r}-\left.\frac{\Lambda^2_
{\gamma}-\mu^2_{\gamma}}{2r}e^{-\Lambda_{\gamma}r}\right].\nonumber\\
\label{s3}
\end{eqnarray}

The quadratic spin-orbit component is neglected (see the discussion in 
the previous subsection).

\section{Parameterization}
\label{sec:parameterization}

In this section we present two parameterizations of our extended GBE CQM. 
One parameterization leaves out spin-orbit forces. The other one takes into 
account all relevant force components produced by the different meson 
exchanges, and thus includes also spin-orbit forces.

\subsection{Extended GBE CQM without spin-orbit forces}
\label{subsec:without_ls}

Usually it is expected that spin-orbit forces play only a minor role 
in hadron spectroscopy. Therefore they are left out in most CQMs. In 
extending the GBE CQM we may, in a first step, try this option too and 
set $V_{\gamma}^{LS}\left(\vec{r}_{ij}\right)=0$ in eqs. \eqref{v1}
and \eqref{s1}.

In the parameterization of the various potential parts the 
constituent quark masses $m_{i}$ are set to the usual values
adopted in CQMs. The meson masses are taken from the 
compilation of the Particle Data Group \cite{PDG}. The mixing of the 
$\eta_0$ and $\eta_8$ mesons, whose mixing angle is $-11.5^\circ$ (as 
determined from the squares of the meson masses),
was neglected in the previous 
pseudoscalar GBE CQM because the mixing effect had been found to be 
unimportant. We maintained this attitude in the parameterization of 
the extended models. The mixing of the vector $\omega_0$ and 
$\omega_8$ mesons, however, is much larger, namely $38.7^\circ$.
Therefore, we took care of this mixing and assumed it to be ideal, i.e. 
with a mixing angle of $35.3^\circ$. No mixing was foreseen for the 
scalar mesons. We remark, however, that all of these assumptions for 
the meson masses are not very relevant. They have to be seen in the 
light of the values of the corresponding cut-off parameters $\Lambda$ 
determined in the fit.

For the quark-meson coupling constants
one may derive suitable estimates from the phenomenologically known 
$\pi$-$N$, $\rho$-$N$, and $\omega$-$N$ coupling constants using the 
Goldberger-Treiman relation. This procedure should lead to reasonable 
magnitudes at least for the coupling constants ${g^2_{ps}}/{4\pi}$ of
pseudoscalar meson exchange as well as the vector and tensor coupling 
constants ${(g^V_{v})^2}/{4\pi}$ and ${(g^T_{v})^2}/{4\pi}$, 
respectively, of vector meson exchange 
(for details see ref. \cite{Gl:02}). For the pseudoscalar coupling
constants we have made the
additional assumption that there is no difference between the 
octet and singlet exchanges, i.e., ${g^2_{ps,8}}/{4\pi} = {g^2_{ps,0}}/{4\pi}$.
For the scalar meson exchange we 
have assumed that the quark-meson coupling constant ${g^2_{s}}/{4\pi}$
is of equal magnitude as in the pseudoscalar case.

The numerical values of all the predetermined parameters are summarized
in table \ref{tab:fix_parameters}. 

\begin{table}[h]
\centering
\caption{Predetermined parameters of the extended GBE CQM (for both 
cases, without and with spin-orbit forces). For additional 
explanations see the text.}
\label{tab:fix_parameters}
\begin{tabular}{lll}
\noalign{\smallskip}\hline\noalign{\smallskip}
\vspace{3mm}
$m_u$            = 340 MeV & $m_d$            = 340   MeV & 
$m_s$          = 507 MeV\\
$\mu_{\pi}$      = 139 MeV & $\mu_{K}$        = 494 MeV & 
$\mu_{\eta}$   = 547 MeV\\
$\mu_{\eta'}$    = 958 MeV & $\mu_{\rho}$     = 770   MeV & 
$\mu_{K^*}$    = 892 MeV\\
$\mu_{\omega_8}$ = 947 MeV & $\mu_{\omega_0}$ = 869   MeV & 
$\mu_{\sigma}$ = 680 MeV\\
\vspace{3mm}
$\mu_{a_0}$      = 980 MeV & $\mu_{\kappa}$   = 980   MeV & 
$\mu_{f_0}$    = 980 MeV\\
${g^2_{ps,8}}/{4\pi}$       =  0.67 & ${(g^V_{v,8})^2}/{4\pi}$ 
= 0.55 & ${(g^V_{v,0})^2}/{4\pi}$ = 1.107 \\
$({g_{ps,0}}/{g_{ps,8}})^2$ =  1    & ${(g^T_{v,8})^2}/{4\pi}$ 
= 0.16 & ${(g^T_{v,0})^2}/{4\pi}$ = 0.0058\\
${g^2_{s}}/{4\pi}$          =  0.67 &\\
\noalign{\smallskip}\hline
\end{tabular}
\end{table}

In addition to the predetermined parameters, the extended GBE CQM 
(without spin-orbit forces) relies on seven fit parameters. Two of 
them concern the confinement interaction: the strength $C$ of the linear 
potential and the constant $V_{0}$ fixing the ground state of the 
spectrum. As indicated above, the value of $C$ is found rather close 
to the magnitude of the string tension of QCD (of approximately 0.1 
GeV$^{2}$). We remark that such a (strong) value for the strength of 
the linear confinement potential is compatible in a CQM only if the 
relativistic expression for the kinetic-energy operator is used. 
Otherwise, in a purely nonrelativistic CQM, this strength would have 
to be chosen unrealistically small \cite{Glozman:1996wq}.

The rest of the free parameters is furnished by the cut offs inherent 
in the meson-exchange potentials. They stem from the finite extension 
of the quark-meson vertices according to eq. \eqref{form_factor}. The
various $\Lambda_{\gamma}$'s are expected to be different for the 
different meson exchanges. Instead of varying them freely, we
prescribed a linear dependence of the cut-off parameters
on the specific meson masses and adjusted only the 
parameters $\Lambda_{\pi}$, $\Lambda_{\rho}$, and $\Lambda_{\sigma}$
occurring therein by a fit to the baryon spectra. Specifically, 
the different scaling prescriptions read:
\vspace{3mm} \newline
Pseudoscalar meson exchange:
\begin{equation}
\Lambda_{\gamma}=\Lambda_{\pi}+\left(\mu_{\gamma}-\mu_{\pi}\right),\qquad 
\gamma=\pi,\:\eta.
\label{pa3}
\end{equation}
%
%
Vector meson exchange:
\begin{equation}
\Lambda_{\gamma}=\Lambda_{\rho}+\left(\mu_{\gamma}-\mu_{\rho}\right),
\qquad \gamma = \rho ,\:K^*,\:\omega_8,\:\omega_0.
\label{pa2}
\end{equation}
%
%
Scalar meson exchange:
\begin{equation}
\Lambda_{\gamma}=\Lambda_{\sigma}+\left(\mu_{\gamma}-\mu_{\sigma}\right),
\qquad \gamma = f_0,\:a_0,\:\kappa,\:\sigma.
\label{pa1}
\end{equation}
Only $\Lambda_{K}$ and $\Lambda_{\eta '}$ are exempted from this 
prescription and are varied independently. This turned out to be 
favorable for an optimal description of all the light and strange 
baryon excitation spectra.

The values of the seven free parameters of the extended GBE CQM (without
spin-orbit forces) are summarized in table \ref{tab:fit_parameters_noLS}.

\begin{table}[h]
\centering
\caption{Free parameters of the extended GBE CQM without spin-orbit forces.}
\label{tab:fit_parameters_noLS}
\begin{tabular}{lll}
\noalign{\smallskip}\hline\noalign{\smallskip}
\vspace{3mm}
$C$  =   1.935 fm$^{-2}$ & V$_0$  =  $-$336 MeV &  \\
$\Lambda_{\pi}$   =  834 MeV    & $\Lambda_{\rho}$  =   1145 MeV  
& $\Lambda_{\sigma}$ =1513 MeV \\
$\Lambda_{K}$  =  1420 MeV   & $\Lambda_{\eta '}$  =  1400 MeV   
&     \\
\noalign{\smallskip}\hline
\end{tabular}
\end{table}

\subsection{Extended GBE CQM with spin-orbit forces}
\label{subsec:with_ls}

In principle, spin-orbit forces are generated by both the confinement 
and the hyperfine interactions. Their net effect should be small, 
however, as 
one does not observe large level splittings from experiments. Still, 
we have aimed at a version of the extended GBE CQM that does include 
spin-orbit forces too. They might be relevant in applications 
beyond spectroscopy.

Instead of employing the explicit expressions for the 
$\vec{L} \cdot \vec{S}$ forces generated by the different meson exchanges 
in eqs. \eqref{v5} and \eqref{s3}, we used a single spin-orbit term 
\begin{eqnarray}
V^{LS}_{\gamma}\left(\vec{r}\:\!\right)&=&-\frac{(g^{LS})^2}{4\pi}\frac{1}
{2m_im_j}\left[\mu^3_{\gamma}\left(\frac{1}{\mu^2_{\gamma}r^2}+\frac{1}
{\mu^3_{\gamma}r^3}\right)e^{-\mu_{\gamma}r}\right.-\nonumber\\[4pt]
&&-\Lambda^3_{\gamma}\left(\frac{1}{\Lambda^2_{\gamma}r^2}+\frac{1}
{\Lambda^3_{\gamma}r^3}\right)e^{-\Lambda_{\gamma}r}-\left.\frac{
\Lambda^2_{\gamma}-\mu^2_{\gamma}}{2r}e^{-\Lambda_{\gamma}r}\right]
\nonumber\\[4pt]
&&(\gamma=\rho, K^{*}, \omega_8, \omega_0, f_0, a_0, \kappa, \sigma)
\label{ls1}
\end{eqnarray}
with a uniform strength ${(g^{LS})^2}/{4\pi}$, which is 
treated as an open parameter.

The assumption of a spin-orbit force as in eq. (\ref{ls1}) is 
motivated by the following findings. The spin-orbit components in the 
hyperfine interaction are a-priori determined from the (vector and 
scalar) meson exchanges. In particular, their strengths are fixed by 
the corresponding quark-meson coupling constants. The spin-orbit 
forces from the confinement, however, remain uncertain in any case,
both with respect to their form and strength. In this regard, one 
cannot escape to assume an ad-hoc spin-orbit contribution in the 
parameterization of the quark-quark interaction. In fact, this problem 
has been studied in the work \cite{Gl:02} by keeping the spin-orbit 
interactions of the meson exchanges from eqs. \eqref{v5} and \eqref{s3}
fixed and varying the strength of the additional spin-orbit force
from the confinement of the form as in eq. \eqref{ls1}. One arrives 
at a certain combination of spin-orbit forces at the cost of 
additional open parameters. It was found that there 
is essentially no difference of such a version of an extended GBE CQM 
from the one presented here. Therefore, in our study of the role of 
spin-orbit forces we contented ourselves with the form as specified 
in eq. (\ref{ls1}). 

All the other interactions are kept the same as in the case of the 
extended GBE CQM of the previous subsection. Also the values of the 
predetermined parameters are maintained as given in table
\ref{tab:fix_parameters}.

\begin{table}[h]
\caption{Free parameters of the extended GBE CQM with spin-orbit forces.}
\label{tab:fit_parameters}
\begin{tabular}{lll}
\noalign{\smallskip}\hline\noalign{\smallskip}
\vspace{3mm}
$C$  =   1.935 fm$^{-2}$ & V$_0$  =  $-$336 MeV &  \\
$\Lambda_{\pi}$   =  834 MeV    & $\Lambda_{\rho}$  =   1145 MeV  
& $\Lambda_{\sigma}$ =1513 MeV \\
\vspace{3mm}
$\Lambda_{K}$  =  1420 MeV   & $\Lambda_{\eta '}$  =  1400 MeV & \\  
${(g^{LS})^2/4\pi}$ = 0.8  \\
\noalign{\smallskip}\hline
\end{tabular}
\end{table}

The extended GBE CQM with spin-orbit forces now relies on eight open 
parameters, two for the confinement and six for the hyperfine 
interaction. While the confinement parameters $C$ and $V_0$
as well as the meson 
cut-offs $\Lambda$ assume the same values as before (in the case 
without spin-orbit forces), one has an additional open parameter, 
namely, the strength ${(g^{LS})^2}/{4\pi}$ of the overall
spin-orbit force \eqref{ls1}. The values of the free parameters
of the extended GBE CQM with spin-orbit forces are summarized in
table \ref{tab:fit_parameters}.

\section{Results}
\label{sec:results}
We are now presenting the spectra of the light and strange baryons 
produced by the two versions of the extended GBE CQM defined in the 
previous sections. For the $N$ and $\Delta$ spectra all resonance
levels are shown up to an excitation energy of about 1800 MeV.
For the strange baryons all analogous octet and decuplet states are
given, as well as the two lowest singlet states in the $\Lambda$
spectrum. The comparison with experiment
is made along the data compiled by the Particle Data Group \cite{PDG};
only three- and four-star resonances with known $J^P$ are considered.

\subsection{Extended GBE CQM without spin-orbit forces}
\label{subsec:results_without_ls}

\begin{figure}[ht]
\begin{center}
\resizebox{0.42\textwidth}{!}{
\includegraphics{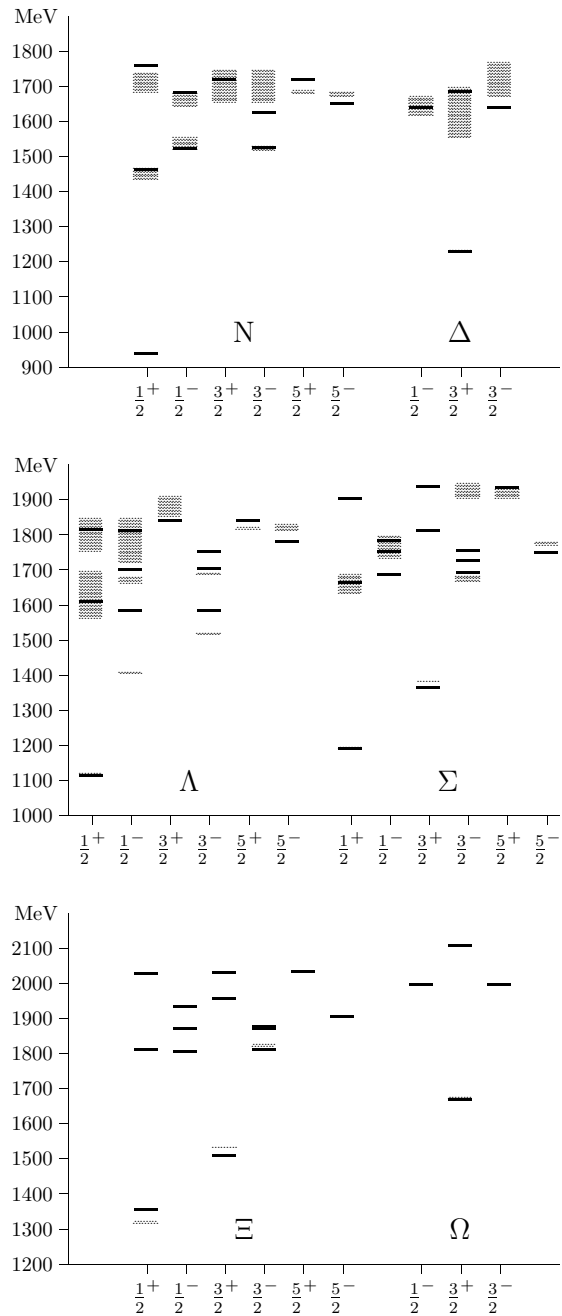}}
\end{center}
\caption{Energy levels of the lowest light and strange baryon states for 
the extended GBE CQM without spin-orbit forces. The shadowed boxes represent
the experimental values and their uncertainties \cite{PDG}.}
\label{fig:ext_GBE_noLS}
\end{figure}

Figure~\ref{fig:ext_GBE_noLS} shows the spectra of the light and strange baryons 
as predicted by the extended GBE CQM without spin-orbit forces (see
section~\ref{subsec:without_ls}). Evidently, the theoretical results 
(solid horizontal bars) are in overall good agreement with the experimental
values (shown as gray boxes). In particular, the 
ground states are described reasonably well. Upon examining the 
$N$ spectrum in more detail, one observes that the 
correct level ordering of positive- and negative-parity excitations is 
achieved in a satisfactory manner; the Roper resonance $N(1440)$ lies 
well below the first negative-parity resonance $N(1535)$. In addition, 
the level splittings within the experimentally almost degenerate 
multiplets remain generally small. In this respect the extended GBE 
CQM resembles the spectral properties of the pseudoscalar GBE CQM 
with spin-spin hyperfine interactions only (cf. figure \ref{fig:ps_gbe}).
However, for the present version of the extended GBE CQM (without spin-orbit 
forces), one also observes deficiencies with regard to the
$\frac52^+ N(1680)$ and $\frac52^- N(1675)$ states; their practical 
degeneracy observed in experiment is not brought about. We require 
explicit spin-orbit forces in order to improve these levels (see the 
next subsection).

The strange baryons are reproduced reasonably well at least with 
respect to the ground states and some 4-star resonances.
Like other CQMs, also the GBE CQM has difficulties in reproducing 
the $\Lambda$ spectrum. While the level ordering of positive- 
and negative-parity states is in principle correct (namely, opposite 
to the $N$ spectrum), the $\Lambda(1405)$ is missed by far. 
Obviously, this state cannot be described as a pure $\{QQQ\}$ state. 
It is strongly influenced by the nearby $K$-$N$ decay threshold.
A similar effect might still influence the $\Lambda(1520)$; it is 
predicted too high by about 60 MeV.

In the PDG listings one finds a $\Xi(1690)$ resonance with 3-star
status, however, with uncertain $J^P$. It is therefore not shown
in our figures. We find no theoretical level that could match such
a state in the relevant energy region. The next excitations beyond
the decuplet ground state $\Xi(1530)$ lie close to 1800 MeV. Here,
one could also think about an influence of the $K$-$\Sigma$
threshold. For instance, for the lowest $\frac12^- \Xi$ a similar 
effect as for the $\frac12^- \Lambda(1405)$ could occur and it could 
relate to the $\Xi$ resonance experimentally seen at an energy 
around 1690 MeV.  

\subsection{Extended GBE CQM with spin-orbit forces}
\label{subsec:results_ls}

\begin{figure}[h]
\begin{center}
\resizebox{0.42\textwidth}{!}{
\includegraphics{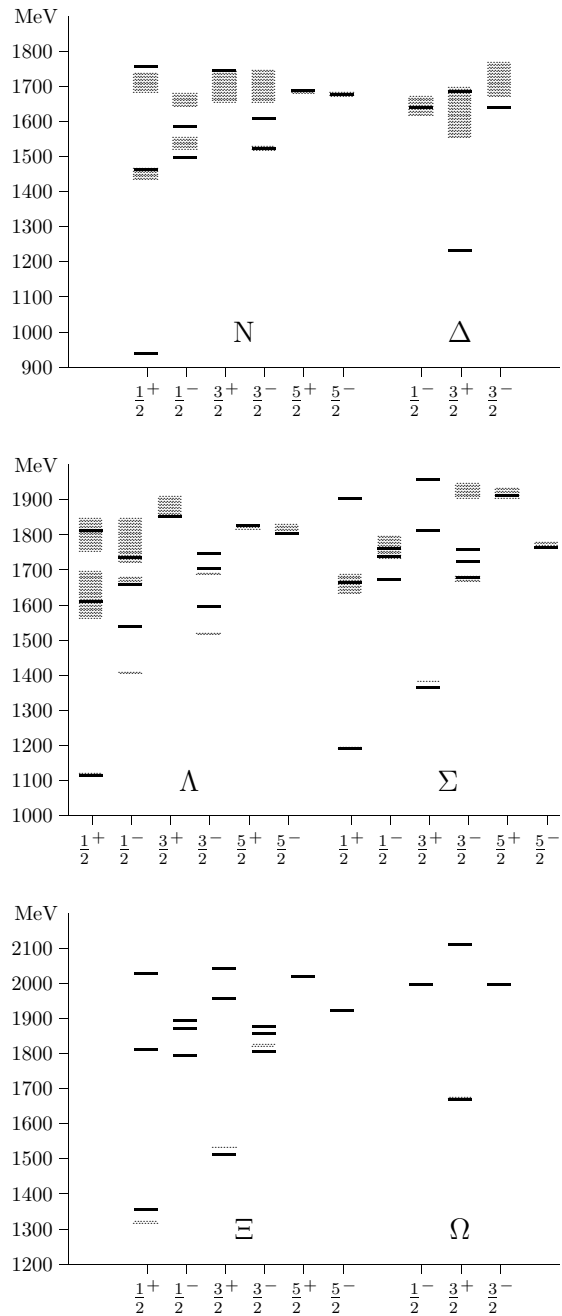}}
\end{center}
\caption{Energy levels of the lowest light and strange baryon states for 
the extended GBE CQM including spin-orbit forces. The shadowed boxes
represent the experimental values and their uncertainties \cite{PDG}.}
\label{fig:ext_GBE_LS}
\end{figure}

The spectra of the extended GBE CQM with spin-orbit forces (see subsection
\ref{subsec:with_ls}) are shown in figure \ref{fig:ext_GBE_LS}. In 
general, the spectral properties of this version are similar to the 
previous one but the almost degenerate levels $\frac52^+ N(1680)$ and
$\frac52^- N(1675)$ in the $N$ spectrum as well as the $\frac52^+ 
\Lambda(1820)$ and $\frac52^- \Lambda(1830)$ in the $\Lambda$ 
spectrum are now improved. The same is true for the
$\frac52^+ \Sigma(1915)$ and $\frac52^- \Sigma(1775)$. The splitting 
of the $\frac52^+ \Xi$ and $\frac52^- \Xi$ gets reduced.
These achievements are due to the 
additional spin-orbit forces. Such a behavior is not obtained in 
neither one of the other versions of the GBE CQM (cf. figures
\ref{fig:ps_gbe} and \ref{fig:ext_GBE_noLS}).

\section{Conclusions}
\label{sec:conclusions}

The GBE CQM by the Graz group \cite{GPVW:98} so far relied on the 
spin-spin part of the pseudoscalar meson exchange only. Here,
we have studied further possibilities offered by GBE dynamics for 
the hyperfine interaction of constituent quarks in light and strange 
baryons. In particular, we have taken into account the effects of 
multiple Goldstone-boson exchange through the exchanges of vector and 
scalar mesons in addition to the pseudoscalar ones. All relevant force 
components have been considered.

We have presented two versions of the extended GBE CQM, one without 
and one with spin-orbit forces. The relativistic quark-model 
Hamiltonian involves only a handful of open parameters.
The hyperfine interaction needs five fit parameters in the case without
spin-orbit forces and six in the case with spin-orbit forces; the
confinement interaction is always the same, namely, a 
linear potential whose strength is congruent with the one deduced
from lattice QCD calculations. In both cases a reasonable 
description of the excitation spectra of all light and strange 
baryons can be achieved. The two versions differ in the quality of 
reproducing higher-lying resonance levels. The almost degeneracy of  
the $\frac52^+ N(1680)$ and $\frac52^- N(1675)$ levels in the $N$
spectrum as well as the $\frac52^+ \Lambda(1820)$ and $\frac52^-
\Lambda(1830)$ levels in the $\Lambda$ spectrum can only be obtained 
in the case when spin-orbit forces are included. At the same time
the description of the $\frac52^+ \Sigma(1915)$ and
$\frac52^- \Sigma(1775)$ splitting is also improved.

The GBE dynamics is now comprehensively included in the quark-model 
Hamiltonian for baryons. This Hamiltonian is equivalent to a covariant
mass operator to be used in further studies within relativistic (i.e. 
Poincar\'e-invariant) quantum mechanics. Its eigenstates are readily 
obtained in the rest frame of any baryon. They can be boosted to an 
arbitrary reference frame in either one of the possible forms of
relativistic quantum mechanics (instant, front, point forms etc.).
Most economically (and accurately) this can be done in point form, 
where the Lorentz transformations remain interaction-free. 

It will be most interesting to repeat the investigations of baryon 
reactions so far done with the pseudoscalar version of the GBE CQM, 
in particular, the calculations of the nucleon electroweak structure 
\cite{WB:01,GR:01,BG:02,Berger:2004yi} and of the mesonic decays
of the baryon resonances \cite{MW:03,Melde:2004xj}.
One may expect new insights into the properties of the corresponding 
observables once all force components (especially the tensor and 
spin-orbit forces) are included. Of course, the extended versions
of the GBE CQM are now also better amenable to further studies, such 
as the $N$-$N$ interaction (cf., e.g., the works by Bartz and Stancu
\cite{BS:99}). The different force components needed in these respects
are directly brought about by the present GBE CQMs.

\end{document}